# Plasma Damage Free *in situ* etching of β-$Ga_2O_3$ using Solid Source Gallium in LPCVD System

Saleh Ahmed Khan[1], Ahmed Ibreljic[1], A F M Anhar Uddin Bhuiyan[1, a)]

[1]*Department of Electrical and Computer Engineering, University of Massachusetts Lowell, MA 01854, USA*

[a)] Corresponding author Email: anhar_bhuiyan@uml.edu

**Abstract**

This work demonstrates an *in situ* etching technique for β-$Ga_2O_3$ using solid-source metallic Gallium (Ga) in a low-pressure chemical vapor deposition (LPCVD) system, enabling clean, anisotropic, plasma damage-free etching. Etching behavior was systematically studied on ($\bar{2}$01) β-$Ga_2O_3$ films and patterned (010) β-$Ga_2O_3$ substrates as a function of temperature (1000-1100 °C), Ar carrier gas flow (80-400 sccm) and Ga source-to-substrate distance (1-5 cm). The process exhibits vapor transport- and surface-reaction-limited behavior, with etch rates reaching a maximum of ~2.25 μm/hr on (010) substrates at 1050 °C and 2 cm spacing. Etch rates decrease sharply with increasing source-to-substrate distance due to reduced Ga vapor availability, while elevated temperatures enhance surface reaction kinetics through increased Ga reactivity and suboxide formation, leading to an enhanced etch rates. In-plane anisotropy studies using radial trench patterns reveal that the (100) orientation produces the most stable etch front, characterized by smooth, vertical sidewalls and minimal lateral etching, consistent with its lowest surface free energy. In contrast, orientations such as (101), which possess higher surface energy, exhibit pronounced lateral etching and micro-faceting. As the trench orientation progressively deviates from (100), lateral etching increases. Facet evolution is observed between (100) and ($\bar{1}$02), where stepped sidewalls composed of alternating (100) and ($\bar{1}$02) segments progressively transition into a single inclined facet, which stabilizes along (100) or ($\bar{1}$02) depending on the trench orientation. The (100)-aligned fins exhibit minimal bottom curvature, while (201)-aligned structures display increased under-etching and trench rounding. Collectively, these findings establish LPCVD based *in situ* etching as a scalable, damage-free, and orientation-selective technique for fabricating high-aspect-ratio β-$Ga_2O_3$ 3D structures in next-generation power devices.

The ultra-wide bandgap β-$Ga_2O_3$ has garnered significant attention for next-generation power and RF electronics due to its ultrawide bandgap (~4.8 eV), high breakdown field strength (~8 MV/cm), controllable n-type doping, and the availability of large-area melt-grown native substrates [1-4]. These properties collectively yield an exceptionally high Baliga's figure of merit, surpassing those of GaN and SiC, and establishing β-$Ga_2O_3$ as a compelling platform for kilovolt-class power conversion, high-frequency amplification, and radiation-hardened electronics [5-12]. Importantly, the ability to grow thick epitaxial layers using scalable techniques such as Metalorganic Chemical Vapor Deposition (MOCVD) [13-18], Molecular Beam Epitaxy (MBE) [19-21], Halide Vapor Phase Epitaxy (HVPE) [22, 23], and LPCVD [24-26] has accelerated the development of β-$Ga_2O_3$ power devices. This progress in β-$Ga_2O_3$ epitaxy have directly enabled the development of high-performance vertical and lateral devices including trench Schottky barrier diodes (SBDs) [27-30], trench MOSFETs [31, 32], and FinFETs [33, 34]. In β-$Ga_2O_3$ vertical devices, where p-type doping is inherently absent, the use of confined and scaled 3D geometries such as vertical fins, mesas, and trenches is essential for controlling current flow and managing electric field distribution. However, the ability to reliably fabricate such high-aspect-ratio 3D structures hinges on achieving controllable, anisotropic, and damage-free etching of β-$Ga_2O_3$. Conventional dry etching approaches, including inductively coupled plasma etching with $BCl_3/Cl_2$, offer vertical profiles but are known to induce substantial plasma damage, including oxygen vacancy formation, stoichiometric deviation, and deep-level trap generation [35, 36]. These effects can significantly degrade interface quality and device performance, lowering breakdown voltages and Schottky barrier heights, while increasing leakage and carrier scattering [37-40]. To mitigate these effects, different wet chemical etching methods using hot phosphoric acid ($H_3PO_4$), hydrofluoric acid (HF), potassium hydroxide (KOH), or tetramethylammonium hydroxide (TMAH), have been explored as plasma-free approaches to minimize surface damage during β-$Ga_2O_3$ processing [41-44]. These methods offer smooth etched surfaces and low defect densities, but are typically isotropic, lack orientation selectivity, and pose challenges for fabricating vertical, high-aspect-ratio structures. Metal-assisted chemical etching (MacEtch) has also been employed to fabricate submicron-scale fin structures [33, 34, 45-47], achieving aspect ratios exceeding 40:1 with smooth sidewalls and low interface trap densities (~$2.7\times10^{11}$ $cm^{-2}eV^{-1}$), which have enabled high-performance devices such as ultra-low-hysteresis FinFETs [33, 34]. However, challenges remain including its relatively low etch rates (~1-4 nm/min), and reduced Schottky barrier heights due to surface oxygen loss and reconstruction [46, 47]. More recently, HCl-based dry etching in HVPE systems enables anisotropic, plasma damage-free etching of β-$Ga_2O_3$, primarily through the formation of volatile gallium chloride species with high vapor pressures [48-53]. In addition, forming gas-based thermal etching in tube furnace systems has also been demonstrated for $Ga_2O_3$ etching [54]. Recently, in situ Ga-assisted etching approaches in MBE and MOCVD chambers, utilizing atomic Ga flux [30, 55, 56] or precursors such as triethylgallium (TEGa)[57] or tert-butyl-chloride (TBCl)[58], have demonstrated facet-selective etching via the formation of volatile gallium suboxide or chloride species, providing valuable insights into the role of crystallographic orientation and precursor chemistry in enabling anisotropic etching. To date, however, *in situ* etching of β-$Ga_2O_3$ with a systematic investigation of its crystallographic directionality and anisotropic behavior using solid-source Ga within a LPCVD platform has not been reported.

In this work, we demonstrate such an approach by employing metallic Ga in a LPCVD system to enable process-compatible, scalable, and damage-free etching of β-$Ga_2O_3$. The use of solid Ga, together with inert gas ambient and plasma-free conditions, eliminates the need for organometallic precursors and complex equipment, offering a low-cost, practical and integration-friendly route for fabricating high-quality β-$Ga_2O_3$ structures. Notably, LPCVD growth system has also been established as a promising platform for developing thick β-$Ga_2O_3$ films, with recent reports demonstrating high growth rates and excellent electronic transport properties [24, 25], including controllable carrier density, promising electron mobility and low background carrier concentration. These combined capabilities position LPCVD as a versatile processing environment for both high-quality growth and anisotropic patterning of β-$Ga_2O_3$.

A schematic of the LPCVD-based *in-situ* etching system used in this study for β-$Ga_2O_3$ surface patterning is illustrated in Figure 1. The process was carried out in a custom-built horizontal LPCVD chamber, where etching is enabled by a thermally driven suboxide reaction between β-$Ga_2O_3$ and solid-source metallic Ga under an oxygen-deficient environment. The overall chemical reaction is expressed as: 4 Ga (s) + $Ga_2O_3$ (s) → 3 $Ga_2O$ (g). This reaction results in the formation of volatile $Ga_2O$, which is continuously removed by the vacuum system. Argon (99.9999% purity) was employed as the carrier and purge gas. High-purity (99.99999%) gallium pellets served as the solid etchant source, placed upstream relative to the substrate holder inside the chamber. The LPCVD chamber was ramped up from room temperature to the target etching temperature at a ramp rate of 10 °C/min, maintained at that temperature for one hour, which was defined as the etching duration, followed by cooling down to room temperature. The etching experiments were performed at temperatures ranging from 1000 °C to 1100 °C, under system pressures of 0.5 to 2.1 Torr corresponding to argon flow rates between 80 and 400 sccm. During the process, both the Ga source and substrate holder were maintained at the same temperature. Two types of samples were investigated: LPCVD-grown ($\bar{2}01$) β-$Ga_2O_3$ films grown on 6° off-axis c-sapphire substrates, as reported in our earlier work [24], and Fe-doped (010) β-$Ga_2O_3$ substrates patterned with $SiO_2$ hard masks. The masks were fabricated by depositing a 100 nm-thick plasma-enhanced chemical vapor deposition (PECVD)-grown $SiO_2$ layer, followed by optical lithography to define window openings. Prior to the etching, all samples were cleaned sequentially with acetone, IPA, and DI water, followed by nitrogen drying. The Ga source-to-substrate distance was systematically varied between 1 and 5 cm to investigate its influence on etch rate and uniformity. During etching, Ga vapor diffuses toward the substrate surface, selectively reacts with exposed β-$Ga_2O_3$ regions and forms volatile $Ga_2O$ suboxide. Regions protected by the $SiO_2$ mask remain unetched, enabling precise pattern transfer. After etching, the $SiO_2$ mask was removed using a diluted buffered oxide etchant (BOE, 1:50), and the etched structures were characterized by field emission scanning electron microscopy (FESEM) (JEOL JSM 7401F) and atomic force microscopy (AFM) (Park XE-100). Etch depths and sidewall profiles were extracted from FESEM cross-sections and AFM surface line scans to determine etch rates and anisotropy.

Figure 2 presents a comprehensive assessment of vertical etch rates in β-$Ga_2O_3$ as a function of Ga source-to-substrate distance, substrate temperature, and argon flow rate. As shown in Figure 2(a), etch rates for ($\bar{2}01$)-oriented samples exhibit a strong dependence on both distance and temperature. At each temperature, the etch rate peaks when the substrate is closest to the Ga source (1 cm) and decreases steadily with increasing distance due to reduced Ga vapor availability and gas-phase transport losses. For any given source-to-substrate distance, the etch rate increases with temperature, indicating the role of thermally activated surface reaction kinetics and enhanced $Ga_2O$ desorption. This temperature-driven increase is most evident in the intermediate spacing regime (2–3 cm), where both Ga delivery and surface reactivity combinedly influence the overall etch rate. However, at the shortest distance (1 cm), where Ga vapor is already abundant, the effect of temperature is less pronounced, suggesting that the surface may be saturated with Ga and that the process is limited by byproduct desorption from the surface or reaction kinetics. Temperature-dependent behavior is further explored in Figures 2(b) and 2(c) for both ($\bar{2}01$) and patterned (010) β-$Ga_2O_3$, confirming consistent etch rate enhancement from 1000 °C to 1050 °C. Notably, the etch rate for (010) β-$Ga_2O_3$ reaches a maximum of ~2.25 μm/hr at 1050 °C, but slightly declines at 1100 °C, possibly due to parasitic gas phase reactions or Ga depletion at the surface at elevated temperature. Similarly, increasing the argon flow rate improves etch rate up to 200 sccm by enhancing Ga vapor transport as shown in Figure 2(d), but further increase to 400 sccm leads to a significant drop, likely due to the Ga dilution and reduced residence time near surface. Figure 2(b) compares the temperature-dependent etch rates of ($\bar{2}01$)-oriented and patterned (010) β-$Ga_2O_3$ at a fixed Ga source-to-substrate distance of 2 cm and Ar flow rates of 200 sccm. Across the entire temperature range, the (010) orientation consistently exhibits higher etch rates than ($\bar{2}01$). This difference is attributed to the crystallographic anisotropy of β-$Ga_2O_3$, where the (010) surface has higher surface energy than ($\bar{2}01$) and greater chemical reactivity toward Ga-induced suboxide formation [59].

Figure 3 investigates the in-plane anisotropy of LPCVD etching on (010) β-$Ga_2O_3$ by evaluating the sidewall profiles of trenches patterned in multiple crystallographic directions. A radial spoke-wheel pattern, shown in Figure 3(a), was used to define trenches along different in-plane directions. The etching was performed at 1050˚C for one hour using Ar flow rate of 200 sccm. Figures 3(b–l) present high-magnification SEM images of the etched sidewalls corresponding to each crystallographic direction after mask removal. These results reveal significant variation in sidewall roughness, verticality, and lateral etch behavior as a function of trench orientation, indicating the anisotropic etch response driven by the monoclinic β-$Ga_2O_3$ crystal structure. Among all orientations, the trench aligned along the (100) sidewall plane [Figure 3(b)] exhibits the smoothest and most vertical sidewalls, with minimal roughness and well-defined (100) faceting. This observation aligns well with density functional theory predictions[59] and previous experimental etching studies [49, 55] that identify the (100) surface as having the lowest surface energy and the highest stability under etching conditions. This low reactivity promotes directional, layer-by-layer removal without excessive lateral undercut, making (100) the most favorable orientation for high-fidelity trench fabrication. In contrast, trenches oriented along planes such as (102), (101), (201) and (301) [Figures 3(i)–3(l)] exhibit pronounced lateral etching and sidewall roughening. Among all orientations, (101) shows the most lateral undercut, indicating it is one of the least stable directions during etching. This behavior is consistent with its highest surface energy and reduced thermodynamic stability as reported by Schewski et al.[59]. Such energetically unfavorable planes are more reactive and less capable of maintaining vertical, smooth sidewalls during anisotropic etching processes. This observation also aligns with experimental findings from HCl-based dry etching in HVPE systems, where the (101) sidewalls on (010) β-$Ga_2O_3$ substrates exhibited the highest lateral etch rates [49]. Intermediate trench orientations, including ($\bar{6}01$), ($\bar{3}01$), ($\bar{2}01$), ($\bar{1}01$), ($\bar{1}02$), and (001) [Figures 3(c–h)], show partial preservation of verticality but with clear evidence of stepped or striped sidewall features. These morphologies arise from the formation of mixed facets and step bunching during the etching, as the system attempts to minimize local surface energy through facet rearrangement, as discussed in more detail in subsequent sections. The (001) sidewall [Figure 3(h)], despite being commonly used for epitaxy, also shows sidewall roughness, likely due to its higher surface energy compared to (100), though less than that of (101) or (102) facets [59]. Overall, the pronounced anisotropy in sidewall morphology across all in-plane directions indicates that the (100) orientation yields the most stable and vertical etch front, while orientation such as (101) is more prone to lateral etching and sidewall roughening due to their higher surface energy and reduced thermodynamic stability, as further corroborated by quantitative analysis of lateral etch rates.

Figure 4 presents a polar plot of lateral etch rates extracted from spoke-patterned trench arrays etched into (010) β-$Ga_2O_3$ substrates at 1000 °C and 1050 °C. The lateral etch rate, defined as the amount of trench broadening beyond the original mask-defined width, is plotted as a function of in-plane crystallographic orientations, offering a quantitative measure of the anisotropy in sidewall etching behavior. The plot reveals that lateral etch rates is strongly orientation dependent. Trenches aligned along the (100) and ($\bar{1}$01) orientations consistently exhibit the lowest and second lowest lateral etching at both temperatures, respectively, indicating highly anisotropic, vertical profiles, correlating with their low surface energies. Similar trends have also been reported for HCl-based etching in HVPE system [49]. In contrast, orientations involving (201), (101), (102) and (001) exhibit much higher lateral etch rates across both temperatures, with (101) demonstrating the most pronounced lateral broadening among all orientations. These orientations expose high-energy surfaces that are more susceptible to isotropic attack and sidewall roughening, consistent with previous reports on HCl-based and MBE etching of β-$Ga_2O_3$, where such planes were found to degrade etch profile fidelity due to increased reactivity [49, 55]. When comparing the two temperatures, 1050 °C etching results in increased lateral etch rates for all orientations relative to 1000 °C. This trend suggests that higher substrate temperatures enhance both Ga transport and surface reaction kinetics, thereby accelerating the formation and desorption of volatile $Ga_2O$ species. While elevated temperatures promote deeper vertical trench propagation, as also evidenced in Figure 2(b) for both $(\bar{2}01)$ and (010) β-$Ga_2O_3$ orientations, they concurrently increase the reactivity of exposed surfaces. As a result, less thermodynamically stable crystal planes become more susceptible to isotropic etching, leading to enhanced lateral broadening and reduced etch anisotropy in those directions.

Figure 5 provides high-resolution SEM images highlighting the evolution of sidewall faceting in trenches etched along progressively rotated in-plane directions on the (010) β-$Ga_2O_3$ surface. While trenches aligned along the (100) orientation maintain smooth, vertical sidewalls with abrupt terminations, other directions such as $(\bar{6}01)$, $(\bar{3}01)$, and $(\bar{2}01)$ exhibit visibly rougher sidewalls and evidence of stepped micro faceting as shown in Figures 5(a–b). A closer inspection of the (100) trench edge [Figure 5(c)] reveals smooth and vertical sidewalls with relatively flat trench bottom, consistent with its low lateral etch rates and anisotropic behavior. However, as the trench direction rotates toward the $(\bar{1}02)$ orientation, a clear transformation in sidewall morphology is observed across Figures 5(d) through 5(h). For instance, the $(\bar{6}01)$-oriented trench [Figure 5(d)] displays a mixed-facet sidewall, composed of alternating (100) and $(\bar{1}02)$ facets, which intersect at approximately 90°, consistent with the crystallographic relationship between these planes in monoclinic β-$Ga_2O_3$. These step-like features suggest a local surface reconstruction during etching, where unstable or higher-energy orientations such as those exposed in off-axis directions evolve into combinations of lower-energy facets to reduce the system's overall surface energy. As the trench orientation rotates away from (100) toward intermediate directions such as $(\bar{3}01)$ and $(\bar{2}01)$ [Figures 5(e) and 5(f)], the sidewalls exhibit progressively rougher morphologies characterized by stepped or scalloped features. Notably, the vertical segments associated with (100) facets become shorter, while the horizontal segments likely corresponding to ($\bar{1}02$) facets increase in length. This shift in facet proportion suggests persistent surface instability and an ongoing transition between competing low-energy planes, reflecting the system's attempt to minimize surface energy in the absence of alignment with a single dominant etch-stable orientation. Interestingly, as the trench alignment approaches ($\bar{1}02$) [Figure 5(h)], the sidewall transitions into a more uniform inclined facet with reduced step formation compared to intermediate orientations, suggesting that ($\bar{1}02$) appears to be a relatively stable etch direction, capable of maintaining uniform surface propagation without decomposing into sub-facets. This behavior is consistent with previous observations during HVPE homoepitaxial growth of (001) β-$Ga_2O_3$, where (100) and ($\bar{1}02$) planes were reported to form stable step terraces due to their lower surface energies [60]. This progressive change is captured schematically in Figure 5(i), which illustrates how the (100) facet initially dominates but is gradually replaced by the inclined facet as the trench orientation rotates. The evolution from mixed-facet to single-facet morphology supports a thermodynamically driven relaxation process, where surface energy minimization governs the final sidewall configuration.

Figure 6 evaluates the impact of crystallographic orientation on the trench geometry and fin profile resulting from LPCVD Ga-based etching, performed at 1050˚C for one hour using Ar flow of 200 sccm. Figures 6(a–d) present tilted SEM images of etched trench arrays aligned along the (100), (001), ($\bar{2}01$), and (201), respectively. While all trenches reach comparable vertical depths (~2.25 μm at 1050 °C), distinct differences in fin width and sidewall morphology are apparent depending on orientation. The (100)-aligned trenches [Figure 6(a)] exhibit the most well-defined geometry, featuring sharp, vertical sidewalls with minimal lateral broadening and the wide fin width (~1.77 μm), in line with its thermodynamically stable etch front observed in Figures 3 and 5. In contrast, fins formed along ($\bar{2}01$) and (201) [Figures 6(c–d)] display significant lateral narrowing (fin widths ~1.25–1.27 μm), reflecting increased under-etching and rougher sidewalls consistent with the facet instabilities previously observed. This orientation-dependent behavior is further supported by the comparative trench profiles shown in Figure 6(e), obtained via AFM line scans across all four directions. The trench with (100) sidewalls exhibits the steepest profile with a relatively flat trench bottom, indicating strong anisotropic vertical etching and minimal lateral deviation. By contrast, trenches aligned with (201) exhibit increased bottom curvature, a signature of enhanced sidewall erosion and reduced etch directionality. Intermediate cases, such as (001) and ($\bar{2}$01), fall between these two, displaying modest undercut and slight trench bottom rounding. This curvature trend mirrors previous findings in Ga-assisted MBE etching studies [55], where bottom rounding was attributed to limited Ga delivery at the trench base and surface reflow during suboxide formation. While AFM measurements may introduce slight convolution effects due to tip geometry in the immediate vicinity of steep sidewalls, the overall trench profile trends, particularly the observed bottom curvature, remain consistent for comparative analysis and are in good agreement with the sidewall morphology observed by FESEM across orientations.

In summary, this study establishes LPCVD technique as a viable and plasma-damage free *in-situ* etching approach for β-$Ga_2O_3$, enabling selective and anisotropic etching through a thermally driven suboxide reaction. Beyond achieving high etch rates and damage-free sidewalls, the work provides a detailed understanding of how orientation, temperature, carrier gas flow and source-to-substrate spacing influence etch kinetics and morphology. The pronounced dependence of sidewall quality on crystallographic direction, captured through lateral etch rates and facet evolution, demonstrates the inherent link between β-$Ga_2O_3$'s monoclinic symmetry and its etching response. Particularly, in-plane anisotropy was systematically mapped, revealing that the (100) orientation with lowest surface energy yielded the steepest, smoothest profiles with minimal bottom curvature, while other planes with higher surface energy exhibits pronounced lateral etching and roughening. The correlation between etch rate, sidewall faceting, and trench geometry provides direct evidence of surface-energy-driven etch behavior across different in-plane directions. These findings not only validate LPCVD-based etching as a process-compatible and scalable alternative to plasma based etching methods for patterning β-$Ga_2O_3$ but also offer critical process structure insights for designing high-aspect-ratio β-$Ga_2O_3$ 3D structures for high-power device architectures.

**Acknowledgement**

The authors acknowledge the funding support from the National Science Foundation (NSF) under award numbers 2501623 and 2532898.

**Data Availability**

The data that support the findings of this study are available from the corresponding author upon reasonable request.

**Conflict of Interest**

The authors have no conflicts to disclose.

**Figure 1**

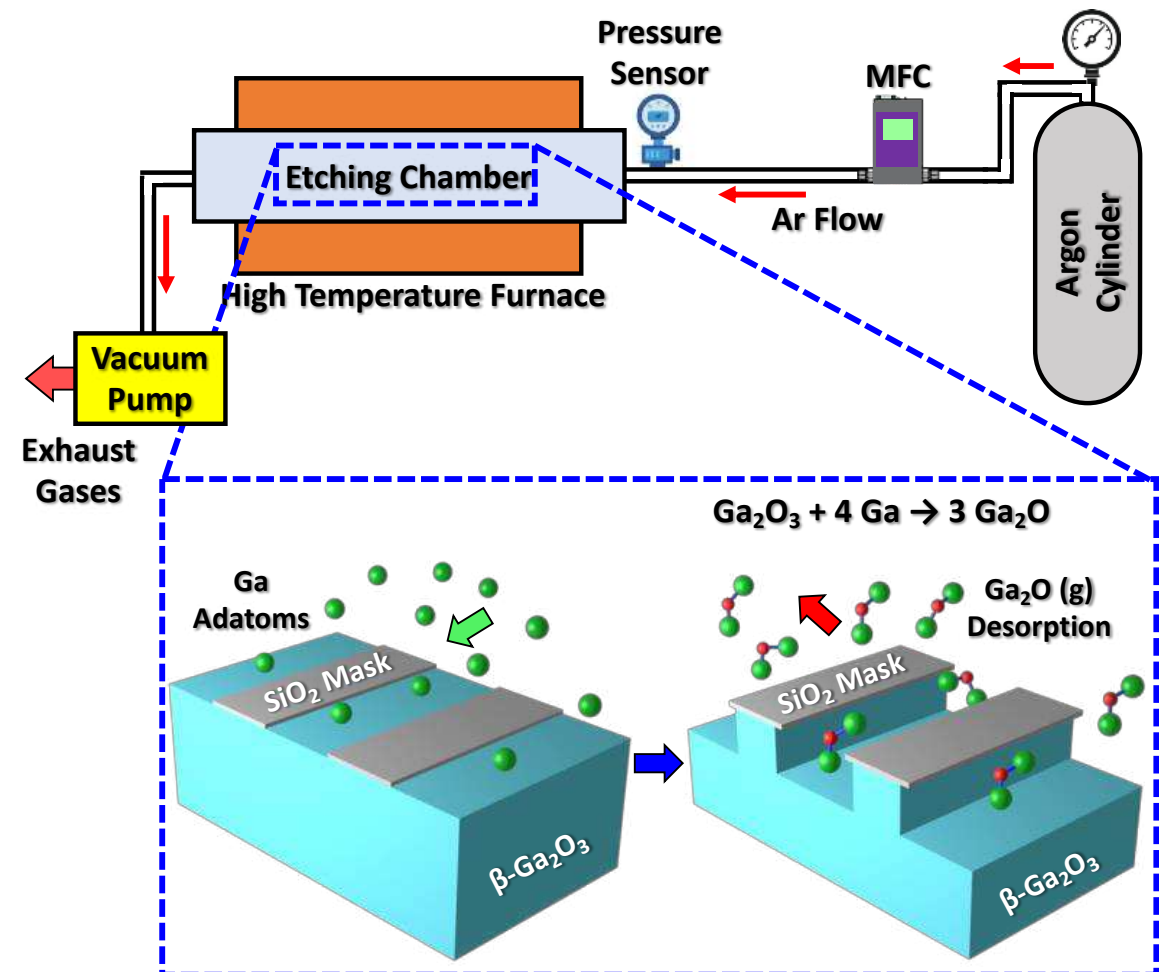

**Figure 1.** Schematic of LPCVD-based in-situ etching of β-$Ga_2O_3$, where Ga adatoms selectively react with exposed β-$Ga_2O_3$ surfaces to form volatile $Ga_2O$ sub-oxide, enabling anisotropic, plasma-free etching through $SiO_2$ mask-defined regions.

**Figure 2**

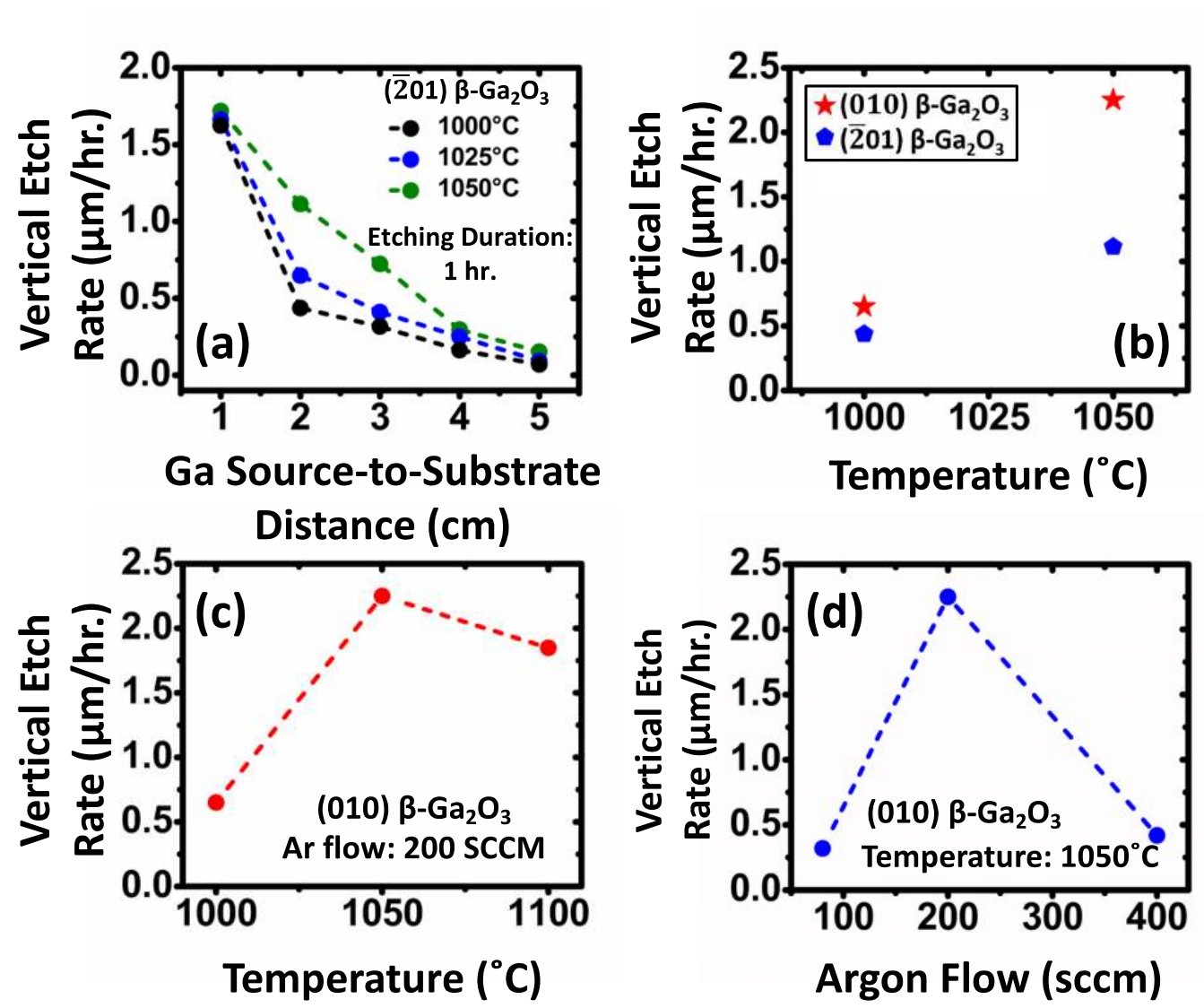

**Figure 2.** Vertical etch rate of $(\bar{2}01)$ β-$Ga_2O_3$ films and patterned (010) β-$Ga_2O_3$ substrates as a function of (a) Ga source-to-substrate distance at various temperatures, (b) substrate temperature for both orientations at fixed distance of 2 cm and Ar flow of 200 sccm, (c) substrate temperature for (010) β-$Ga_2O_3$ at fixed distance of 2 cm and Ar flow of 200 sccm, and (d) Ar carrier gas flow for (010) β-$Ga_2O_3$ at a distance of 2 cm and temperature of 1050 °C.

**Figure 3**

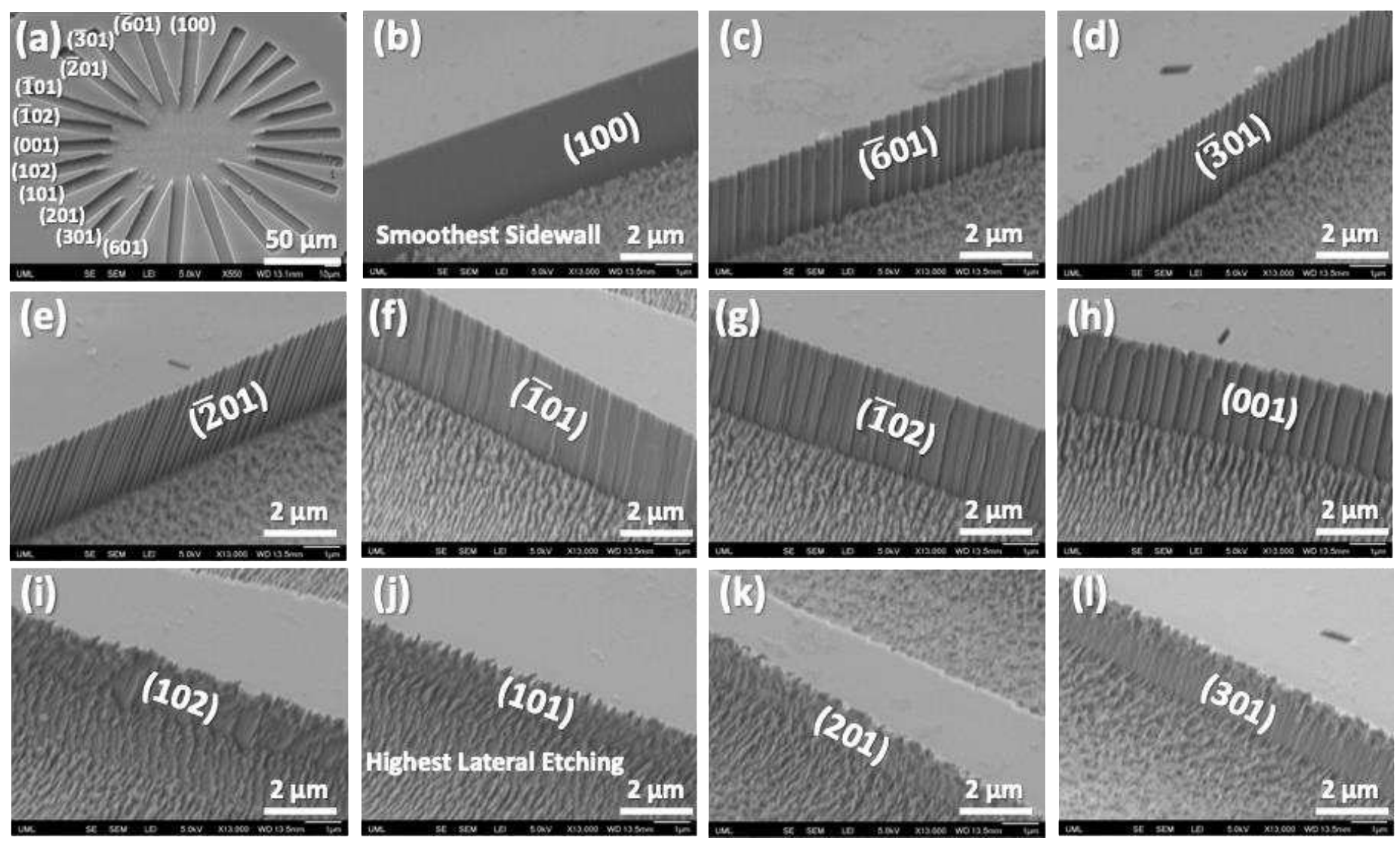

**Figure 3.** (a) Tilted top-view FESEM image (~60° tilt) of a radial spoke-wheel trench pattern etched into a (010) β-$Ga_2O_3$ substrate, with trenches aligned along various in-plane crystallographic orientations. (b-l) Corresponding high-magnification sidewall SEM images showing orientation-dependent variations in sidewall morphology.

**Figure 4**

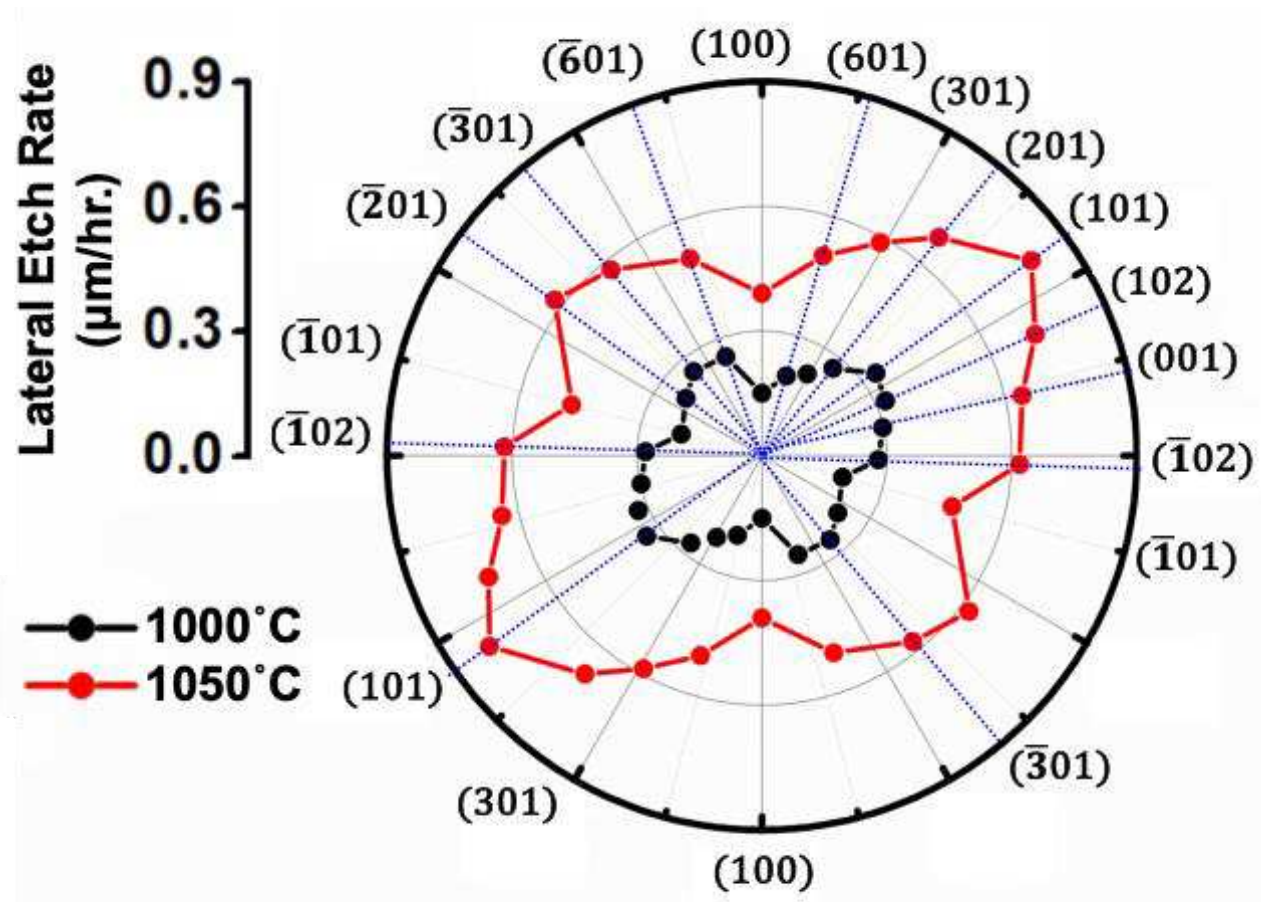

**Figure 4.** Polar plot of lateral etch rates measured on (010) β-$Ga_2O_3$ patterned substrates etched at 1000 °C and 1050 °C for one hour. The plot reveals pronounced in-plane anisotropy, with the (100) orientation exhibiting the lowest lateral etch rates, indicative of highly anisotropic and stable etching behavior. In contrast, the (101) orientation consistently shows the highest lateral etch rates across both temperatures, reflecting its higher surface energy and lower thermodynamic stability during etching.

**Figure 5**

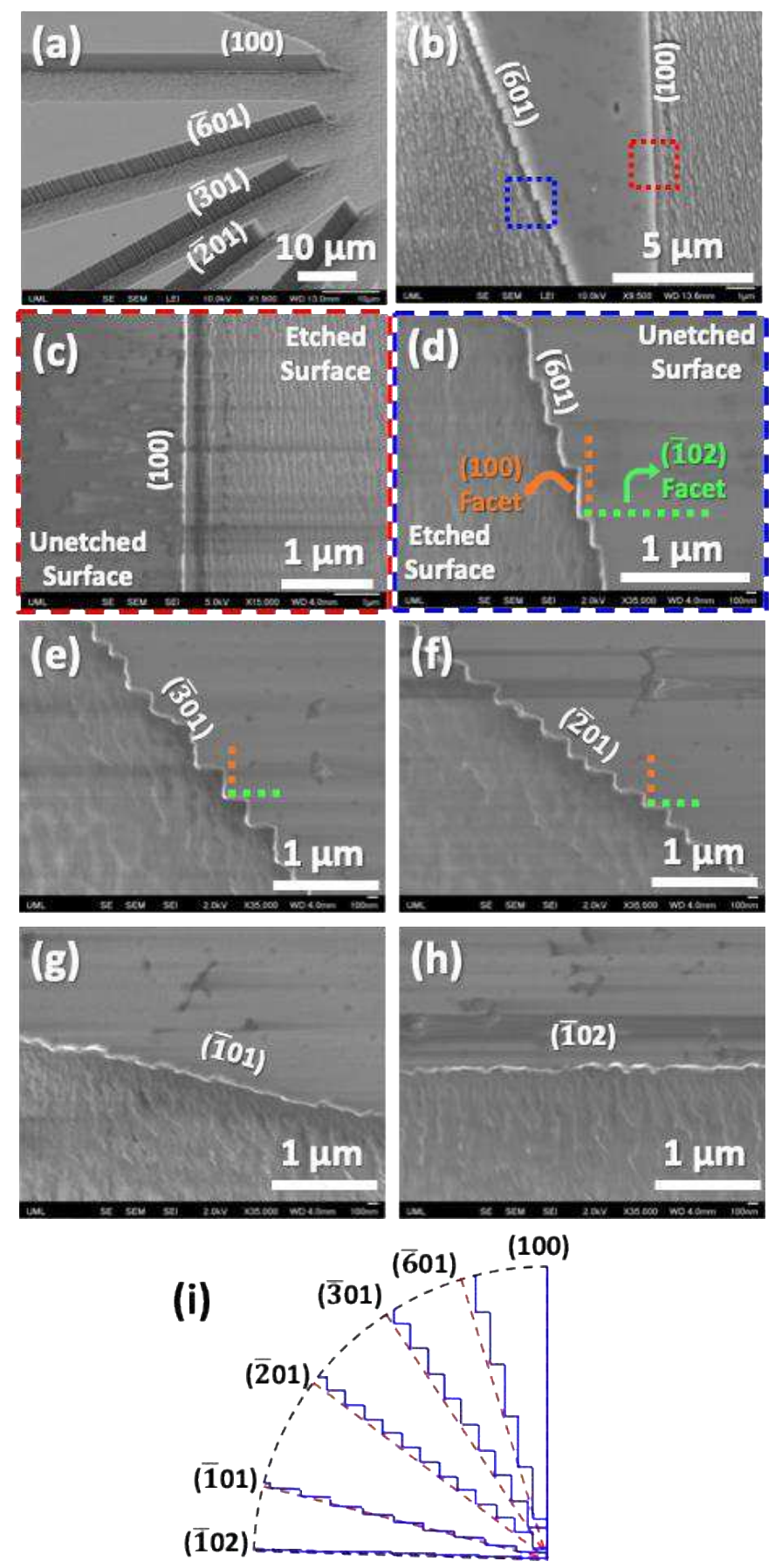

**Figure 5.** (a) Tilted FESEM image showing progressive sidewall roughness on (010) β-$Ga_2O_3$ trenches oriented at increasing angles from the (100) plane. (b) Top-view SEM highlighting the difference between the smooth sidewall along (100) and the stepped sidewall observed along the $(\bar{6}01)$ plane. (c–h) High-magnification SEM images showing stepped facet evolution, with segments composed of alternating (100) and $(\bar{1}02)$ facets. (i) Schematic representation of the sidewall geometry illustrating the angular transition from vertical (100) to inclined $(\bar{1}02)$ facets as the trench orientation rotates away from (100).

**Figure 6**

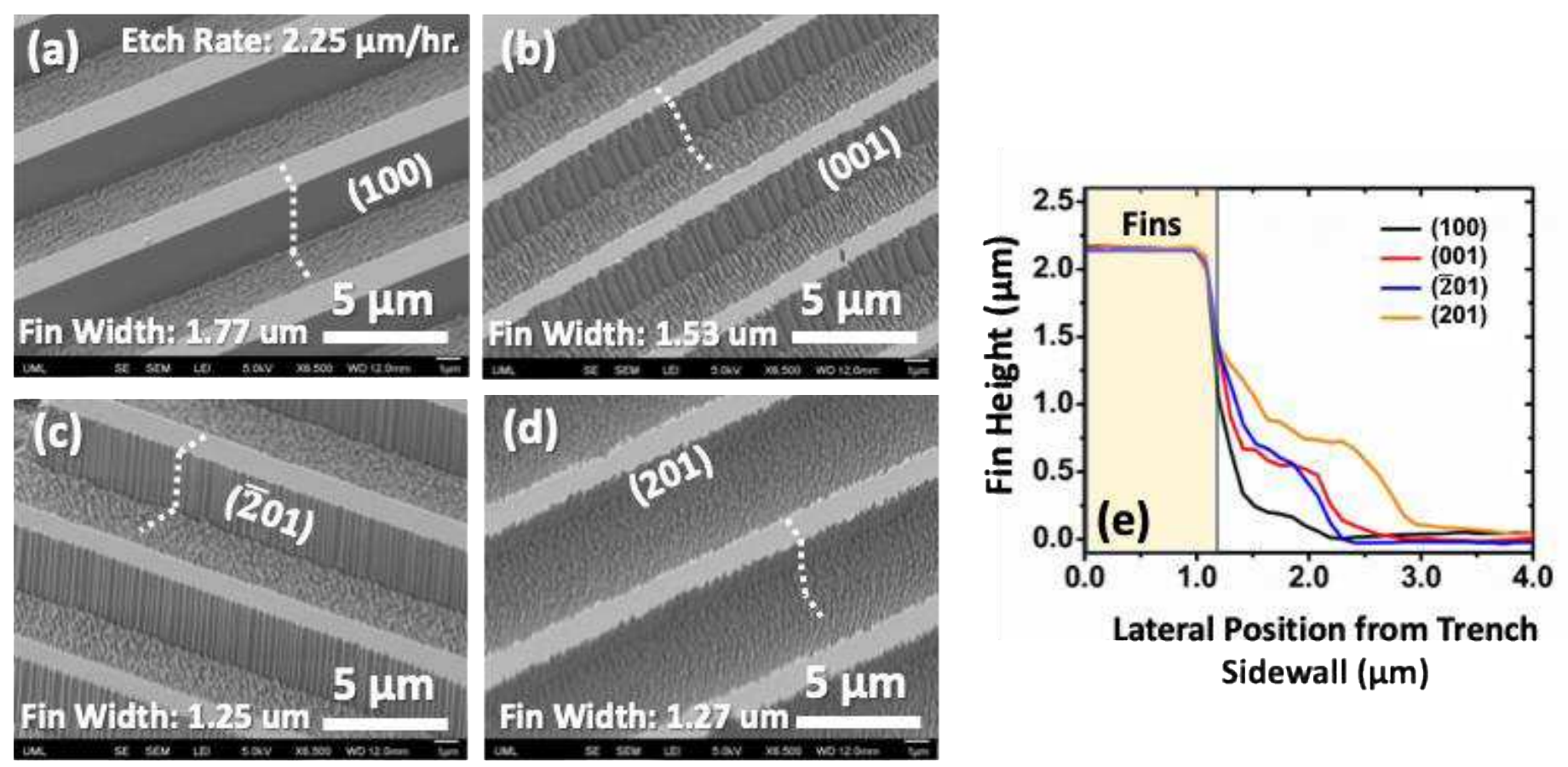

**Figure 6.** (a-d) Tilted FESEM images of etched trench arrays along different crystallographic orientations on (010) β-$Ga_2O_3$, showing variation in fin width and sidewall morphology. (e) AFM line scans comparing trench profiles across different planes, highlighting differences in fin top sharpness and the extent of trench bottom curvature based on sidewall orientation.